\documentstyle[11pt]{article}

\newcommand{\Bbb}{\bf}
\def\frak#1{{#1}}

\def\bra#1{\langle #1 |}	
\def\ket#1{| #1\rangle}		

\def\be{\begin{equation}}
\def\en{\end{equation}}
\def\bea{\begin{eqnarray}}
\def\ena{\end{eqnarray}}
\def\bean{\begin{eqnarray*}}
\def\enan{\end{eqnarray*}}

\newcommand{\gsl}{{\frak {sl}}}
\newcommand{\ggl}{{\frak {gl}}}

\newcommand{\slh}{\widehat{\gsl}}
\newcommand{\glh}{\widehat{\ggl}}
\def\slth{\slh_2}

\def\glth{\glh_2}                      %

\def\uq{U_q(\slth)}

\def\uqglt{U_q(\glth)}

\def\ch{{\rm ch}\,}
\def\sh{{\rm sh}\,}

\def\qdet{\hbox{$q$-det}}
\def\ket#1{|#1\rangle}

\def\Z{{\Bbb Z}}
\def\Q{{\Bbb Q}}

\def\R{{\Bbb R}}

\def\hL{\widehat{L}}

\def\bR{\overline{R}}

\begin{document}

\title{ Relativistic~Calogero-Sutherland~Model :
Spin~Generalization, Quantum~Affine~Symmetry and
Dynamical~Correlation~Functions}

\author{Hitoshi Konno${}^*$\cr
{\it  Yukawa Institute for Theoretical Physics,}\cr
{\it Kyoto University, Kyoto 606, Japan.}\cr
{\rm ( YITP/K-1118, hep-th/9508016 ) }}
\date{August 4, 1995}

\maketitle
\begin{abstract}
Spin generalization of the relativistic Calogero-Sutherland model
is constructed by using the affine Hecke algebra and shown to
possess the quantum affine symmetry $\uqglt$.
The spin-less model is exactly diagonalized by means of
the Macdonald symmetric polynomials.
The dynamical density-density correlation function
as well as one-particle Green function are evaluated exactly.
We also investigate the finite-size scaling
of the model and show that
the low-energy behavior is described by the $C=1$ Gaussian theory.
The results indicate that the excitations obey the
fractional  exclusion statistics and exhibit the Tomonaga-Luttinger
liquid behavior as well.
\end{abstract}

Recently, Yangian symmetry has been extensively
studied\cite{HHTBP,BGHP,BPS}
in the relation  to
the Calogero-Sutherland model (CSM)\cite{Cal},
the Haldane-Shastry model (HSM)\cite{Hal} as well as conformal
field theory (CFT).
Especially, it is remarkable that  a new structure of
CFT  called spinon
structure has been understood based on
this symmetry\cite{BPS}.
This motived the author in \cite{JKKMP} to
analyze the analogous structure of the level-1
integrable highest weight modules of the quantum affine algebra
$\uq$. These modules are known to have a deep connection with
integrable spin chain models\cite{JM}.
The level-0 action of $\uq$ in the level-1 modules
plays the same role as the Yangian
in CFT. Namely the level-1 modules are completely reducible with
respect to the level-0 action.
However, no physical models related to this level-0 symmetry
have been discussed.

One of the purpose of this Letter is to
propose a model having this symmetry.
We consider the trigonometric limit of the Ruijsenaars-Schneider
model\cite{Ruijs},
which may be considered as a relativistic extension of CSM.
Let $\theta_j\ (j=1,2,..,N)$
be the rapidity variables and $x_j$ be their canonically conjugate
variables.
We impose the canonical commutation relations
$[x_j, \theta_l]=i\delta_{j,l}$
with $\hbar=1$ and use the representation $\theta_l=-i\partial/\partial{x_l}$.
The model is described by the following Hamiltonian $H$ and
momentum operator $P$
\be
H=\frac{c^2}{2}(H_{-1}+H_1),
\qquad P=\frac{c}{2}(H_{-1}-H_1)
\label{hammom}
\en
with $N-$independent integrals of
motion $H_k$ (or $ H_{-k}$) $(k=1,2,..,N)$
\bea
&&H_{\pm k}=\sum_{I\subset \{1,2,..,N \}\atop |I|=k}
\prod_{i\in I\atop j\not\in I}
\Bigl(\frac{{\rm sin}\frac{\alpha}{2}(x_i-x_j\mp i g/c)}
{{\rm sin}\frac{\alpha}{2}(x_i-x_j)}\Bigr)^{1/2}
e^{1/c\sum_{i\in I}\theta_i}\nonumber\\
&&\qquad\qquad\qquad\times\prod_{l\in I\atop m\not\in I}
\Bigl(\frac{{\rm sin}\frac{\alpha}{2}(x_m-x_l\mp i g/c)}
{{\rm sin}\frac{\alpha}{2}(x_m-x_l)}\Bigr)^{1/2}.
\label{ham}
\ena
Here $c$ being the speed of light and $\alpha\in \R_{>0},\ g\in \Q$.
We normalize the mass equal to one.
The model possesses the Lorentz boost generator
$B=-\frac{1}{c}\sum_{i=1}^N x_i$,
and is  Poincar${\acute{\rm e}}$ invariant in the sense
that the operators $H,\ P$, and $B$
satisfy the Poincar${\acute{\rm e}}$ algebra:
\be
[H,P]=0,\qquad [H,B]=iP,\qquad [P,B]=i\frac{H}{c^2}.
\label{poincare}
\en

In the non-relativistic limit $c\rightarrow \infty$,
we recover the Hamiltonian of CSM as
\bea
&&\lim (H-Nc^2)=-\sum_{j=1}^N\frac{1}{2}\Bigl(\frac{\partial}{\partial
x_j}\Bigr)^2+\frac{g(g-1)}{4}\sum_{1\leq j<k\leq N}
\frac{\alpha^2}{\sin^2\frac{\alpha}{2}(x_j-x_k)}\nonumber
\ena
with identification
$\alpha = 2\pi/L$, where $L$ is the length of a ring on which particles
are confined.

\def\pth#1{{p^{\vartheta_{#1}}}}

It is also known that the integrals of motions $H_{k}$
can be gauge transformed to the Macdonald operators\cite{Macbk}.
Let us define new parameters
$p=e^{-\alpha/c}$, $t=p^g$
and new variables
$ \quad z_j=e^{i\alpha x_j}$,
$p^{\pm\vartheta_j}=
e^{\mp \frac{\alpha}{c} z_j\frac{\partial}{\partial{z_j}}}$.
Notice the relation $p^{\pm\vartheta_j}z_j=p^{\pm 1}z_j$.
Then, by using the function
\be
\Delta=\prod_{j,k=1\atop j\not=k}^N
\frac{(z_j/z_k;p)_{\infty}}{(tz_j/z_k;p)_{\infty}},
\en
one has \cite{Diejen}
\be
\Delta^{-1/2}H_{\pm k}\Delta^{1/2}=t^{\mp k(N-1)/2}
D_k(p^{\pm1},t^{\pm 1}).
\label{dhdd}
\en
Here  $D_k(p,t)$ are the Macdonald operators  defined by\cite{Macbk}
\be
D_k(p,t)=t^{k(k-1)/2}\sum_{I\subset \{1,2,..,N  \}\atop |I|=k}
\prod_{i\in I\atop j\in\not\in 	I}\frac{tz_i-z_j}{z_i-z_j}\prod_{i\in
I}\pth{i}.
\nonumber
\en

Now let us discuss a spin generalization of the model
and clarify its quantum affine symmetry.
The model we will consider
is essentially the trigonometric model discussed  by
Bernard et.al.\cite{BGHP}. But it has never connected to the
relativistic CSM.
Let us consider the trigonometric solution $\bR(z)$\cite{JM} of
the Yang-Baxter equation and the operator
$L_{0i}(z)\ (i=1,2,..,N)$ defined by
\be
L_{0i}(z)=\frac{1-q^2z}{(1-z)q}\bR_{0i}(z)=
\frac{zS^{-1}_{0i}-S_{0i}}{1-z}P_{0i}.
\label{lops}
\en
with $P_{ij}(v_i\otimes v_j)=v_j\otimes v_i $ and
\be
S(z)=
\left(\matrix{
-q^{-1}  &         &          &    \cr
&q-q^{-1}&-1& \cr
&-1&0& \cr
 &          &          & -q^{-1} \cr}\right).
\label{Shecke}
\end{equation}
With $V_j\  ( j=0,1,..,N)$ being two dimensional vector spaces,
$\bar L_{0i}(z)$ is regarded as
a linear operator on $V_0\otimes V_i$.
Note also that the operators $S_{jj+1}\ (j=1,2,..,N-1)$  satisfy the
Hecke algebra relations.
\begin{eqnarray}
&&S_{jj+1}-S_{jj+1}^{-1}=q-q^{-1},
\nonumber
\\
&&S_{jj+1}\ S_{kk+1}=S_{kk+1}\ S_{jj+1}
\qquad |j-k|>1,
\label{eqn:Hecke2}\\
&&S_{jj+1}\ S_{j+1j+2}\ S_{jj+1}=S_{j+1j+2}\ S_{jj+1}\ S_{j+1j+2}.
\nonumber
\end{eqnarray}
Define the monodromy matrix $L_0(z)$ by
\bea
&&L_0(z)=L_{01}(z)L_{02}(z)\cdots L_{0N}(z).
\ena
Then the operators $\bR(z)$ and $L_0(z)$ satisfy the relation
\be
\bR_{00'}(z/z')L_0(z)L_{0'}(z')
=L_{0'}(z')L_0(z)\bR_{00'}(z/z').
\label{rllllr}
\en
We use this relation to realize the quantum affine symmetry
$\uqglt$ as well as to define an integrable spin generalization of the model.
For this purpose,
we introduce
the affine Hecke algebra $\hat{H}_N(q)$\cite{BGHP}.
The algebra $\hat{H}_N(q)$ is
generated by $g_{jj+1}\ (j=1,2,..,N-1)$ and $y_j\ (j=1,2,..,N)$
with the relations (\ref{eqn:Hecke2}) for
$g_{jj+1}$ and
\begin{eqnarray}
&&y_j\ y_k=y_k\ y_j, \nonumber\\
&&g_{jj+1}\  y_j\  g_{jj+1}= y_{j+1},\\
&&[g_{jj+1}, y_k]=0, \qquad (j, j+1\not=k).
\nonumber
\end{eqnarray}
We use the following representation of $\hat{H}_N(q)$\cite{JKKMP}.
\begin{eqnarray*}
&&g_{jk}^{\pm 1}={qz_j-q^{-1}z_k\over z_j-z_k}(1-K_{jk})-q^{\mp 1}\\
&&y_j=r_{jj+1}^{-1}\cdots r_{jN}^{-1}\
p^{\vartheta_j}\ r_{1j}\cdots r_{j-1j}
\end{eqnarray*}
with $K_{jk}f(...,z_j,..,z_k,...)=f(...,z_k,..,z_j,...)$ and
$r_{jk}=K_{jk}g_{jk}$.

Since the operators ${y}_j$ ($j=1,\ldots,N$) commute
with each other, the 'quantized' monodromy matrix\cite{BGHP}
 \be
\hL_0(z)=L_{01}(z{y}^{}_1)\cdots
L_{0N}(z{y}^{}_N)
\label{eqn:Lop}
\en
also satisfies the
relation (\ref{rllllr}).
Consider the formal expansion of $\hL_0(z)$ in $z^{\pm1}$ and define
\be
\hL^{\pm}_0(z)=\sum_{\pm n\geq 0}z^{n}
\left(\matrix{
l^{\pm}_{11}[n]  &l^{\pm}_{12}[n] \cr
l^{\pm}_{21}[n]&l^{\pm}_{22}[n] \cr}\right).
\label{explop}
\en
{}From (\ref{lops}) and (\ref{rllllr}),
 we have the relations $l^{+}_{21}[0]=l^{-}_{12}[0]=0$ and
$l^{+}_{jj}[0]l^{-}_{jj}[0]=1$  $\ (j=1,2)$ as well as
\bea
&& \bR_{0{0'}}(z/z')\hL^\pm_0(z)\hL^\pm_{0'}(z')
=\hL^\pm_{0'}(z')\hL^\pm_0(z)\bR_{0{0'}}(z/z'),\label{rlla}\\
&& \bR_{0{0'}}(z/z')\hL^+_0(z)\hL^-_{0'}(z')
=\hL^-_{0'}(z')\hL^+_0(z)\bR_{0{0'}}(z/z').\label{rllb}
\end{eqnarray}

Now let ${\cal F}_N$ be  the space of vectors
$v\in \{f(z_1,z_2,..,z_N)\otimes V^{\otimes N}\}$ satisfying
\be
(g_{jj+1}-S_{jj+1})v=0 \qquad j=1,2,..,N-1.
\label{gsrel}
\en
The  relations (\ref{rlla})  and (\ref{rllb}) define a
level-0 action ${\uqglt}_0$ on ${\cal F}_N$\cite{JKKMP}.
{}From (\ref{eqn:Lop}) and (\ref{explop}), we obtain the following
realization:
\begin{eqnarray*}
&&\pi^{(N)}(e_0)=\sum_{j=1}^N y^{-1}_j q^{h_1}\otimes\cdots\otimes
q^{h_1}\otimes \stackrel{j}{\check{f_1}}\otimes q^{h_2}\otimes\cdots
\otimes q^{h_2} ,\\
&&\pi^{(N)}(f_0)=\sum_{j=1}^N  y_j
q^{-h_2}\otimes\cdots\otimes
q^{-h_2}\otimes \stackrel{j}{\check{e}_1}\otimes q^{-h_1}\otimes\cdots
\otimes  q^{-h_1},\\
&&\pi^{(N)}(e_1)=\sum_{j=1}^N q^{h_2}\otimes\cdots\otimes
q^{h_2}\otimes \stackrel{j}{\check{e}_1}\otimes q^{h_1}\otimes\cdots
\otimes q^{h_1} ,\\
&&\pi^{(N)}(f_1)=\sum_{j=1}^N
q^{-h_1}\otimes\cdots\otimes
q^{-h_1}\otimes \stackrel{j}{\check{f_1}}\otimes q^{-h_2}\otimes\cdots
\otimes  q^{-h_2},\\
&&\pi^{(N)}(q^{\pm h_j})=q^{\pm h_j}\otimes\cdots \otimes q^{\pm h_j} \qquad
j=1,2 ,
\end{eqnarray*}
where $e_1=\pmatrix{0&1\cr0&0\cr}$, $ f_1=\pmatrix{0&0\cr1&0\cr}$,
$h_1=\pmatrix{1&0\cr 0&0\cr}$ and $ h_2=\pmatrix{0&0\cr 0&1\cr}$.

The center of ${\uqglt}_0$ is given by
the quantum determinant ${\qdet}\hL_0(z)$.
Direct calculation shows
\bea
&&{\qdet}\hL_0(z)=q^N\prod_{j=1}^N
\frac{(1-q^{-1}y_jz^{-1})}{(1-q y_jz^{-1})}.
\ena
Expanding ${\qdet}\hL_0(z)$ in the power of $z^{-1}$, one gets the
commuting family of $N-$independent operators
\be
\sum_{i_1<\cdots<i_k}y_{i_1}\cdots y_{i_k} \quad
(k=1,2,..,N).
\label{intmot}
\en

Now we define a model on ${\cal F}_N$ by the
following Hamiltonian  $\hat h$ and  momentum operator $  \hat p$
\be
\hat{h}=\frac{c^2}{2}\sum_{j=1}^N(y^{-1}_j+y_j),\quad
\hat{p}=\frac{c}{2}\sum_{j=1}^N(y^{-1}_j-y_j).
\en
Defining also the operator
$\hat{b}=-\frac{i}{\alpha}\sum_{j=1}^N\ln z_j$,
one can easily show that
$\hat h$, $  \hat p$ and $\hat{b}$
satisfy the Poincar${\rm{\acute e}}$
algebra  (\ref{poincare}).
Furthermore,
in the spin-less sector
of ${\cal F}_N$, for example $\{ f_{sym}(z_1,..,z_N)\otimes
v_+\otimes\cdots\otimes  v_+\}$ with $f_{sym}$ being symmetric
functions, $\hat h$, $\hat p$ as well as all the integrals of motion
(\ref{intmot})
of the model coincide with those of  the relativistic Calogero-Sutherland
model (\ref{hammom})$\sim$(\ref{ham}).
This is due to the following formula\cite{JKKMP} valid on this sector.
\bean
&&D_k(p^{\pm1},t^{\pm1})=(-t^{1/2})^{\pm k(N-1)}
\sum_{i_1<\cdots<i_k}y^{\pm1}_{i_1}\cdots y^{\pm1}_{i_k},
\enan
where we made identification $t=q^2$.
{}From (\ref{dhdd}), this implies
$H=\Delta^{1/2}\ \hat{h}\ \Delta^{-1/2}$ and
$P=\Delta^{1/2}\ \hat{p}\ \Delta^{-1/2}$.
We hence have obtained the integrable spin generalization of the relativistic
Calogero-Sutherland model and shown that it possesses
the quantum affine symmetry ${\uqglt}_0$.

We next  consider the
diagonalization of the spin-less model and
evaluate the dynamical correlation functions.
The diagonalization of the integrals of motion (\ref{ham}) can be
carried out by the Macdonald symmetric polynomials. Let
$\lambda=(\lambda_1,\cdots,\lambda_N)$,
$\lambda_1\geq\cdots\geq \lambda_N\geq 0$ be a partition and
denote the Macdonald symmetric polynomial by $P_{\lambda}(z;p,t)$.
Then one has\cite{Macbk},
\bean
&&D_k(p^{\pm1},t^{\pm1})P_{\lambda}(z;p,t)=\Bigl(\sum_{i_1<\cdots<i_k}
\prod_{l=1}^kt^{N-i_l}p^{\lambda_l}\Bigr)P_{\lambda}(z;p,t).
\enan
Therefore, from (\ref{dhdd}), we obtain the exact eigen values of
$H$ and $P$ as
\bea
&&E_N(\lambda)=c^2\sum_{j=1}^N {\rm ch}\frac{\theta_j}{c},\quad
P_N(\lambda)=c\sum_{j=1}^N {\rm sh}\frac{\theta_j}{c}\label{spectrum}\\
&&\theta_j=\frac{2\pi}{L}\Bigl\lbrace\lambda_j+g\Bigl(\frac{N+1}{2}-j\Bigr)
\Bigr\rbrace,
\label{psrap}
\ena
where we set $\alpha=2\pi/L$.
The corresponding  eigen functions are given by
\be
\Psi_{\lambda}(z)=\Delta^{1/2}P_{\lambda}(z;p,t).
\en
The model thus can be regarded as an ideal gas of $N-$relativistic
pseudo-particles with the pseudo-rapidities (\ref{psrap}).
One should note that the formula (\ref{psrap}) obey the following
Bethe ansatz like equations
\be
L\theta_j=2\pi I_j+
\pi(g-1)\sum_{l=1}^N{\rm sgn}(\theta_j-\theta_l),
\label{tbae}
\en
with $I_j=\lambda_j+\frac{N+1}{2}-j$.

The ground state is given by the function $\Psi_{\phi}(z)=\Delta^{1/2}$
corresponding to the empty partition $\lambda=\phi$.
The  ground state momentum and energy eigenvalues are evaluated as
$P_N^{(0)}=0$ and
\be
E_N^{(0)}
=c^2\sh\frac{\pi gN}{cL}/\sh\frac{\pi g}{cL}.
\label{grenergy}
\en
Hence the ground state can be described as a filled Fermi sea with
pseudo-momenta $P_j^{(0)}={\rm sh}\frac{\theta_j}{c}$
with $-\theta_F\leq \theta_j\leq \theta_F$ $(j=1,2,..,N)$, where
$\theta_F={\pi g}({N-1})/L$.

The dynamical density-density correlation functions
as well as one-particle Green function can be evaluated
by making use of
the Macdonald symmetric polynomial.
We here summarize the results.
To each partition $\lambda$,
we assign a Young diagram ${\cal D}(\lambda)=\{(i,j)|1\leq i\leq
l(\lambda),\ 1\leq j\leq \lambda_i,\quad i,j \in \Z_{>0}\}$.
Let $\lambda'$ be the conjugate partition of $\lambda$.
For each cell $\gamma=(i,j)$ of ${\cal D}(\lambda)$,
we define the quantities
$a(\gamma)=\lambda_i-j$, $a'(\gamma)=j-1$,
$l(\gamma)=\lambda'_j-i$ and $l'(\gamma)=i-1$.
Then we have
\bea
\bra{0}\rho(\xi,t)\rho(0,0)\ket{0}&=&\frac{A_N}{L^2}\sum_{\lambda}
\frac{(1-p^{|\lambda|})^2(\chi^{\lambda}(p,t))^2}
{h_{\lambda}(p,t)h_{\lambda'}(t,p)}
{\cal N}(\lambda)\nonumber\\
&&\qquad\qquad\times
{\rm cos}({\cal P}(\lambda)\xi)e^{-i{\cal E}(\lambda)t},
\label{dddcf}\\
\bra{0}\Psi^{\dagger}(\xi,t)\Psi(0,0)\ket{0}
&=&\frac{A_{N+1}}{A_N}\sum_{\lambda}
\frac{t^{2|\lambda|}\Bigl((t^{-1})_{\lambda}^{(p,t)}\Bigr)^2}
{h_{\lambda}(p,t)h_{\lambda'}(t,p)}
{\cal N}(\lambda)\nonumber\\
&&\qquad\qquad\times e^{-i({\cal E}(\lambda)t-{\cal P}(\lambda)\xi)},
\label{greenf}
\ena
with $\xi$ being a real coordinate conjugate to $P$,
$|\lambda|=\sum\lambda_j$,
${\cal E}(\lambda)=E_N(\lambda)-E_N^{(0)}$,
${\cal P}(\lambda)=P_N(\lambda)$ and
\bean
&&A_N=\prod_{j=1}^N\frac{(pt^{j-1};p)_{\infty}(t;p)_{\infty}}
{(t^{j};p)_{\infty}(p;p)_{\infty}},\\
&& h_{\lambda}(p,t)=\prod_{\gamma\in \lambda}\Bigl(
1-p^{a(\gamma)}t^{l(\gamma)+1}\Bigr),\qquad
 h_{\lambda'}(t,p)=\prod_{\gamma\in \lambda}\Bigl(
1-p^{a(\gamma)+1}t^{l(\gamma)}\Bigr),\\
&&{\cal N}(\lambda)=\prod_{\gamma\in \lambda}
\frac{1-p^{a'(\gamma)}t^{N-l'(\gamma)}}{1-p^{a'(\gamma)+1}t^{N-l'(\gamma)-1}},
\\
&&\chi^{\lambda}(p,t)=\prod_{\gamma\in \lambda\atop \gamma\not=(1,1)}
\Bigl(t^{l'(\gamma)}-p^{a'(\gamma)}\Bigr), \qquad
(a)_{\lambda}^{(p,t)}=\prod_{\gamma\in \lambda}
\Bigl(t^{l'(\gamma)}-p^{a'(\gamma)}a\Bigr).
\enan
For the rational coupling $g=r/s$,
one should remark that the factor $\chi^{\lambda}(p,t)$
(resp.$(t^{-1})_{\lambda}^{(p,t)}$ )
vanishes if the diagram ${\cal D}(\lambda)$ contains the lattice
point $(s+1,r+1)$(resp.$(s,r+1)$).
According to the same argument as by Ha in CSM\cite{Ha},
this indicates that in the thermodynamic limit, only the states  which
contains minimal $r$ quasi-hole excitations accompanied by
$s$(resp. $s-1$) quasi-particles
can contribute as the intermediate states in (\ref{dddcf})
(resp.(\ref{greenf})).
One can thus conclude that
the excitations of the model obey the fractional exclusion
statistics ${\rm \grave{a}}$ la Haldane\cite{Haldane} as in CSM\cite{Ha}.

Furthermore,
the exact spectrums (\ref{spectrum}) allows one to analyze the finite-size
scaling of the model in the thermodynamic limit, $N$, $L$$\rightarrow
\infty$ with $N/L=n$ fixed.
First of all, from (\ref{grenergy}) we obtain the finite-size
correction to the ground state energy as
\be
\lim E_N^0=L\varepsilon_0-\frac{\pi v }{6L}g+O(\frac{1}{L^2}),
\label{grenfss}
\en
where $\varepsilon_0=\frac{c^3}{\pi g}\sh\frac{\pi g n}{c}$ and
$v=c\ \sh\frac{\pi g n}{c}$ are the
ground state energy density
and the velocity of the elementary excitation, respectively.
In comparison
with the general theory\cite{BCN},
one may suspect that the central charge is given by $g$.
However this is not the correct identification\cite{KaYa}.
The central charge should be identified with 1. This can be justified by
calculating the low temperature expansion of the free energy from
(\ref{tbae}). Instead, we here justify it by
deriving the whole conformal dimensions associated with the elementary
excitations. These can be obtained by evaluating the differences
of the total energy and momentum from the groung state eigenvalues
under the change of the particle number (by $\Delta N$) and the
transfer of the $\Delta D-$particles from the left Fermi point to the
right one\cite{KaYa}.
 We hence obtain
the finite-size corrections
\bea
&&\Delta E=\mu\Delta N+
\frac{2\pi v}{L}\Bigl[\ \frac{g}{4}
\Delta N^2+\frac{1}{g}\Bigl(\Delta D+\frac{\Phi}{2\pi}\Bigr)^2\Bigr],
\nonumber\\
&&\Delta P=2 p_F\Delta D+ \frac{2\pi \ch\frac{\pi g n}{c}}{L}
\Delta N\Bigl(\Delta D+\frac{\Phi}{2\pi}\Bigr),\nonumber
\ena
where $\mu=c^2\ch\frac{\pi gn}{c}$ and $p_F=v/g$ are
the chemical potential and  the Fermi momentum, respectively.
We here modified the argument by Kawakami and Yang by considering the
flux excitations $\Phi$ associated with the change of the particle number
$\Delta N$\cite{Haldtani,Ha}. Adding the contribution from the
quasi-particle ($N^+$) and quasi-hole($N^-$)
excitations,
we finally obtain the right and left conformal dimensions
$h^\pm$ as follows.
\be
h^\pm(\Delta N; \Delta D; N^\pm)=\frac{1}{2}\Bigr[
\frac{\sqrt{g}\Delta N}{2}\pm\frac{1}{\sqrt{g}}
\Bigl(\Delta D+\frac{\Phi}{2\pi}\Bigr)\Bigl]^2+N^\pm.
\label{confdim}
\en
Remarkablly, the result does not depend on $c$.
Note  that the flux carried by a particle is $\pi g$ as in
CSM\cite{Haldtani} so that $\Phi=\pi g \Delta N$.
One can thus write (\ref{confdim}) as
\bean
&&h^+=\frac{1}{2g}(\Delta D+g\Delta N)^2 +N^+,\
h^-=\frac{1}{2g}\Delta D^2 +N^-.
\enan
This result indicates that
the effect of the flux excitation is equivalent to
impose the new selection rule $\Delta D=\frac{g}{2}\Delta N$ (mod 1)
on (\ref{confdim}) without $\Phi/2\pi$.
Notice that this selection rule can be obtained from the periodicity
of the plane wave $\exp(i\theta_jx_j)$ under the change $x_j\to x_j+L$.
Hence $h^{\pm}$ with $N^{\pm}=0$ can be regarded as the conformal
dimensions of
the $U(1)-$primary fields in the $C=1$ Gaussian theory.
{}From the results (\ref{dddcf}) and (\ref{greenf}),
we also have succeeded to
obtain  the thermodynamic limit of the dynamical correlation
functions and  their asymptotic form.
The critical exponents thus obtained agree
with Ha's results\cite{Ha} as well as
those obtained from $h^{\pm}$ with assignment $\Delta N=0$ for
the density correlation and $\Delta N=1$ for one-particle Green function.
One can thus conclude
that the model possesses the Tomonaga-Luttinger liquid
property\cite{Hald}.

In the case with the special coupling
$g=2$, the Gaussian theory is known to become the level-1 $su(2)$
Wess-Zumino-Witten theory. This feature is consistent
with the results in Ref.\cite{JKKMP}, where the setting
$t=p^2$, is inevitable to define a new level-0 action of
$\uq$.

In comparison with CSM, our model possesses one extra paprmeter $c$.
The ultra relativistic limit $c\to 0$ is especially interesting.
There one has a decoupling of the left and right-movers.
In addition, the limit $g\to 0$ with $g/c$ fixed reduces
the Macdonald polynomial to
the Hall-Littlewood function\cite{Macbk}. This suggests that a certain
mathematical structre remains in this limit\cite{SKAO}.
The detailed investigation  will be discussed elsewhere.

The author would like to  thank
P.J.~Forrester, T.~Fukui, M.~Jimbo, N.~Kawakami, V.B.~Kuznetsov,
T.~Miwa, K.~Ueno and T.~Yamamoto
for valuable discussions.

\noindent
${}^*$ Yukawa Fellow.
{}

\end{document}